# Magnetocrystalline Anisotropy in a Single Crystal of $CeNiGe_2$

M. H. Jung*, N. Harrison, and A. H. Lacerda

*National High Magnetic Field Laboratory – Pulse Facility, Los Alamos National Laboratory, MS E536 Los Alamos, NM 87545*

P. G. Pagliuso[+], J. L. Sarrao, and J. D. Thompson

*Condensed Matter and Thermal Physics, Los Alamos National Laboratory, MS K764 Los Alamos, NM 87545*

We report measurements on single crystals of orthorhombic $CeNiGe_2$, which is found to exhibit highly anisotropic magnetic and transport properties. The magnetization ratio $M(H//b)/M(H\perp b)$ at 2 K is observed to be about 18 at 4 T and the electrical resistivity ratio $\rho_{//b}/\rho_{\perp b}$ is about 70 at room temperature. It is confirmed that $CeNiGe_2$ undergoes two-step antiferromagnetic transition at 4 and 3 K, as reported for polycrystalline samples. The application of magnetic field along the $b$ axis (the easy magnetization axis) stabilizes a ferromagnetic correlation between the Ce ions and enhances the hopping of carriers. This results in large negative magnetoresistance along the $b$ axis.

---

\* *To whom correspondence should be addressed. E-mail: mhjung@lanl.gov*
 *and also Physics Department, New Mexico State University, NM 88003*
[+] *Present address: Institute of Physics, Uniersity of Campinas, SP 13083, Brazil*

## I. INTRODUCTION

Members of Ce$TX_2$ ($T$ = transition metals and $X$ = semimetallic elements) compounds have been of much interest because of the variety of ground states, such as heavy-fermion behavior in CePtSi$_2$,[1] valence fluctuating in CeRhSi$_2$,[2] heavy-fermion ferromagnetic in CeRuSi$_2$,[3] and nonmagnetic valence fluctuating in CeNiSi$_2$.[4] The nature of the ground state depends on the strength of hybridization and the crystal chemistry of both the transition metal and semimetallic elements. So far, Ce$T$Si$_2$ compounds have been intensively investigated, while Ce$T$Ge$_2$ compounds have not. Nevertheless, one could expect that substitution of semimetallic silicon by the more metallic germanium should cause significant changes in their ground states. To exploit this potential, we have chosen the Ce$T$Ge$_2$ series, where different crystal structures are observed depending on the transition elements. The orthorhombic YIrGe$_2$-type structure is found when $T$ is a 5$d$ transition metal,[5] while the CeNiSi$_2$-type structure is found when $T$ is a 3$d$ transition metal.[6] This paper focuses on results obtained for CeNiGe$_2$ single crystals.

CeNiGe$_2$ crystallizes in the orthorhombic CeNiSi$_2$-type layered structure,[6] which is constructed from deformed fragments of the CeGa$_2$Al$_2$ and $\alpha$-ThSi$_2$ structures. For polycrystalline samples of CeNiGe$_2$, heavy-fermion behavior is found in the specific heat, but with somewhat different values of the Sommerfeld coefficient $\gamma$ = 97 and 220 mJ/K$^2$mol.[4,7] A double peak structure in the electrical resistivity characteristic of Kondo effect under the crystalline field is observed.[7] Magnetic susceptibility and specific heat measurements indicate two antiferromagnetic transitions at $T_N^I$ = 4 K and $T_N^{II}$ = 3 K.[4] However, the nature of two-step antiferromagnetic ordering below 4 K has not yet been elucidated. Furthermore, the anisotropic behavior of such a material with multiple magnetic transitions warrants an investigation with single-crystal samples to be made.

## II. EXPERIMENTAL DETAILS

Single crystals of CeNiGe$_2$ were grown using a Sn-flux method, from which we obtained many platelet single crystals of approximate size $1 \times 3 \times 3$ mm$^3$ with the short length along the $b$ axis. Powder x-ray diffraction pattern revealed them to crystallize in the orthorhombic CeNiSi$_2$-type (space group Cmca) structure. The lattice parameters were $a$ = 4.25(3) Å, $b$ = 16.78(9) Å, and $c$ = 4.21(0) Å, close to those reported previously.[6] The in-plane resistivity and Hall coefficient were measured on one of the platelets using a standard four-probe AC method. The $b$-axis resistivity was estimated from the measurements using a modified four-probe configuration, adapted for layered compounds.[8] Magnetization and specific heat were taken using a Quantum Design SQUID magnetometer and a Quantum Design PPMS system, respectively. Measurements of high-field magnetization to 50 T and magnetoresistance to 20 T were carried out at the National High Magnetic Field Laboratory, Los Alamos.

## III. RESULTS AND DISCUSSION

### A. High anisotropy in magnetic and transport properties

Figure 1 displays the anisotropic behavior of the magnetization $M(H)$ of CeNiGe$_2$, which was measured at 2 and 0.5 K in static field and pulsed field to 4 and 50 T, respectively, applied along the $b$ axis ($H//b$) and in the plane ($H \perp b$). The absence of hysteresis and remanence in $M(H)$ indicates the antiferromagnetic nature of the ground state. There is a very large anisotropy with an easy magnetization direction along the $b$ axis. The magnetization $M(H//b)$ saturates rapidly to a value of 1.2 $\mu_B$/Ce at 4 T and then reaches 1.9 $\mu_B$/Ce at 50 T, but is still smaller than the full moment (2.4 $\mu_B$/Ce) expected for a free $Ce^{3+}$ ion. On the other hand, $M(H\perp b)$ increases linearly with increasing magnetic field. From the inset to Fig. 1, the magnetization ratio of $M(H//b)/M(H\perp b)$ at 2 K is estimated to be about 18 at 4 T. This large anisotropy in response to the magnetic field is also evident in the magnetic susceptibility.

The inverse magnetic susceptibility $\chi^{-1}(T)$ in a field of 0.1 T parallel ($H//b$) and perpendicular to the $b$ axis ($H\perp b$) as a function of temperature from 2 to 350 K is shown in Fig. 2. As expected from the $M(H)$ observations, the $b$-axis magnetic susceptibility $\chi_{//b}$ is much larger than the in-plane magnetic susceptibility $\chi_{\perp b}$ at all temperatures. Above 100 K, the data obey the Curie-Weiss law, $\chi = C / (T - \theta_P)$ with the paramagnetic Curie temperatures of $\theta_{//b} = 31.9$ K and $\theta_{\perp b} = -168.2$ K. This might indicate the development of ferromagnetic and antiferromagnetic exchange interactions for $H//b$ and $H\perp b$, respectively. From the value of $\theta_P = (\theta_{//b} + 2\theta_{\perp b}) / 3$, we can calculate the Kondo temperature $T_K \sim |\theta_P / 2|$,[9] obtaining $T_K = 50$ K. This value is comparable to that ($T_K = 45$ K) determined from the Sommerfeld coefficient ($\gamma = 97$ mJ/K$^2$mol) measured polycrystalline samples[4] and $T_K = 40$ K from $\gamma = 108$ mJ/K$^2$mol for single crystals (which will be discussed in the next section). The high-temperature slopes of $\chi^{-1}$ yield the effective magnetic moments of $\mu_{//b} = 2.11$ $\mu_B$ and $\mu_{\perp b} = 2.22$ $\mu_B$, which are slightly smaller than the value (2.52 $\mu_B$) reported for polycrystalline samples.[4] This implies that the magnetic moments of Ce ions are well localized in this material. The deviation from the Curie-Weiss behavior below 100 K could be attributed to the crystalline-electric-field (CEF) effect. The low-temperature $\chi_{//b}$ and $\chi_{\perp b}$ data exhibit two anomalies at 4 and 3 K indicating antiferromagnetic orderings, as shown in the insets of Fig. 2. The antiferromagnetic transition at 4 K in $\chi_{//b}$ is the most pronounced. The increase of $\chi_{\perp b}$ between 4 and 3 K is possibly due to the incommensurate nature of the antiferromagnetic ordering in the plane.

We have further studied the anisotropic transport properties of $CeNiGe_2$. Figure 3 shows the electrical resistivity $\rho(T)$ with the current parallel ($I//b$) and perpendicular ($I \perp b$) to the $b$ axis. The most striking feature is that the $b$-axis resistivity $\rho_{//b}(T)$ is much larger than the in-plane resistivity $\rho_{\perp b}(T)$. The resistivity ratio $\rho_{//b}/\rho_{\perp b}$ is estimated to be about 70 at room temperature. This anisotropic transport indicates a dominant motion of conduction electrons in the plane, reflecting the anisotropic crystal structure. In the inset of Fig. 3, we have plotted the in-plane resistivity $\rho_{\perp b}(T)$ as a function of the logarithm of the temperature. It is characterized by a broad peak around 80 K, followed by a minimum at ~ 20 K and a steep drop below 4 K. The peak structure is typical of that expected for materials, in which CEF effects influence the Kondo effect.[10] The steep drop below 4 K could be attributed to the combined effect of the reduction of spin-disorder scattering and the development of coherence, as found in other Kondo antiferromagnetic systems.[11] The difference between $\rho_{//b}(T)$ and $\rho_{\perp b}(T)$ is likely to result from different types of hybridization between the localized $4f$ electrons and the conduction electrons and/or anisotropic magnetic ordering between the $b$ axis and in the plane.

**B. Two magnetic phase transitions at low temperatures**

Figure 4 shows the low-temperature data of specific heat $C(T)$, temperature derivative of the in-plane resistivity $d\rho_{\perp b}/dT$ and Hall coefficient $R_H(T)$. Two phase transitions are observed that apparently correspond to two antiferromagnetic transitions at $T_N^I = 4$ K and $T_N^{II} = 3$ K, consistent with previous results on polycrystalline samples.[4, 7]

In Fig. 4(a), the specific heat $C(T)$ exhibits two peaks at 4 and 3 K due to these magnetic phase transitions. From the data at temperatures between 13 and 22 K, the linear Sommerfeld coefficient is estimated to be $\gamma \simeq 108$ mJ/mol K$^2$. From this $\gamma$ value, the Kondo temperature is estimated to be about $T_K = 40$ K,[12] close to that ($T_K = 50$ K) obtained from the paramagnetic Curie temperature $\theta_P$. Even in the magnetically ordered state (0.3 K < $T$ < 1 K), $C(T)$ is well represented by $C(T) = \gamma T + \beta T^3$ with $\gamma \simeq 260$ mJ/mol K$^2$. This suggests that strongly correlated electronic states coexist with magnetic order in this system. Figure 4(b) shows the temperature derivative of in-plane resistivity $d\rho_{\perp b}/dT$. The transition temperatures can be taken to be 4 and 3 K, where $C(T)$ exhibits two peaks, as the sudden-increase points in the $d\rho_{\perp b}/dT$ curve. The coexistence of strongly correlated electronic states with magnetic order might weaken the anomalies at the transition temperatures in $\rho(T)$. It is noteworthy in Fig. 4(c) that $R_H(T)$ is positive at temperatures down to 2 K, indicating that hole-type carriers are dominant in the transport. Furthermore, $R_H(T)$ shows two distinct anomalies, a rapid rise below 4 K and a sharp peak at 3 K. This anomalous Hall effect observed at the magnetic transition temperatures may be attributed to skew scattering involving spin-orbit coupling and spin-flip scattering. However, this term yields values much smaller than the experimental data. Recently, a very large anomalous Hall effect was found in colossal magnetoresistance manganites[13] and was explained by Berry phase theory based on carrier hopping in a nontrivial spin background.[14] Such a Berry phase can change the motion of carriers in the presence of magnetic ordering, giving a large change of the Hall coefficient at the magnetic ordering temperature.

## C. Magnetic field effect on the two transitions

Since the magnetic field can influence the arrangement of magnetic spins, it is interesting to investigate the field effect on the magnetic transitions. We have measured the magnetization $M(H)$ at various temperatures and the specific heat $C(T)$ in applied magnetic fields. Figure 5 shows the isothermal magnetization $M(H)$ of CeNiGe$_2$ at 0.5, 3.5, 5, and 20 K below, between, and above the two transitions ($T_N^I$ = 4 K and $T_N^{II}$ = 3 K) in the field range $0 < H < 10$ T. The magnetization $M(H//b)$ at 0.5 and 3.5 K below $T_N^I$ shows a metamagnetic-like transition around 0.7 T, where d$M$/d$H$ has a maximum. This implies that the antiferromagnetic ground state below $T_N^I$ changes into a field-induced ferromagnetic state along the field direction parallel to the $b$ axis. At 5 K just above $T_N^I$, $M(H//b)$ has no metamagnetic transition but still increases rapidly with magnetic field, whereas $M(H//b)$ at 20 K far above $T_N^I$ increases linearly. These results in $M(H//b)$ are compared with the linear increase of $M(H\perp b)$ at 0.5 K.

The specific heat $C(T)$ was measured at various magnetic fields, zero, 2, 4, and 6 T for $H\perp b$ and $H//b$. As seen in Fig. 6, $C(T)$ in zero field exhibits two peaks at 4 and 3 K due to the magnetic phase transition. In a field of 2 T for $H\perp b$, the upper peak at 4 K shifts to 3.7 K and the lower peak at 2.7 K is unchanged. When the field exceeds 4 T, only one peak is observed at 3 and 2.5 K for $H$ = 4 and 6 T, respectively. It is likely that two peaks at $H \leq 2$ T become combined to form a single peak at $H \geq 4$ T. This field effect is more drastic for $H//b$, the easy magnetization axis: $C(T)$ in a field of 6 T for $H\perp b$ is almost identical to $C(T)$ in 2 T for $H//b$. In other words, the value of magnetic field where two peaks merge into one peak is much lower when the field is applied along the easy axis (Fig. 9). For fields ($H//b$) higher than 4 T, the peak in $C(T)$ disappears and instead develops to a broad maximum. With increasing magnetic field, the maximum shifts towards higher temperatures, reminiscent of a Schottky-type anomaly due to magnetic defects.[15] The entropy $R\ln 2$ is conserved at around 20 K both in zero-field and 6 T, especially for $H//b$.

### D. Anisotropic magnetoresistance and phase diagram

In order to tentatively elucidate the nature of magnetic transitions, we have studied the effects of magnetic field on the resistivity. Figure 7 shows the temperature dependence of electrical resistivity $\rho(T)$ for $I//b$ and $I\perp b$ at constant magnetic fields to 18 T for $H//b$ and $H\perp b$. As one can see, the field effects are quite different between $\rho_{//b}(T)$ and $\rho_{\perp b}(T)$. In the configuration $I//b$ (the upper panels of Fig. 7), the kink at $T_N^I = 4$ K in zero field due to the magnetic transitions disappears in fields $H \geq 2$ T for $H//b$, whereas it remains in fields up to 5 T for $H\perp b$. In the configuration $I\perp b$ (the lower panels of Fig. 7), the kink at $T_N^I$ is suppressed with increasing magnetic field for both $H//b$ and $H\perp b$, although the change of $\rho_{\perp b}(T)$ for $H//b$ is much stronger than that for $H\perp b$. These results correspond to the anisotropy observed in the magnetization measurements; $\Delta M(H//b) > \Delta M(H\perp b)$.

In Fig. 8, the normalized magnetoresistance $\Delta\rho/\rho_o = [\rho(H)-\rho(0)]/\rho(0)$ is plotted as a function of the applied magnetic field at fixed temperatures. With increasing magnetic field for $H//b$ (the left panels of Fig. 8), the magnetoresistance at 2 K is initially positive and then turns negative at higher magnetic fields, giving rise to maxima at 0.5 and 3.2 T for $\rho_{//b}(H//b)$ and $\rho_{\perp b}(H//b)$, respectively. These field values coincide with those in changing the slope of $M(H//b)$. Thus, we could attribute the change of magnetoresistance to the arrangement of magnetic spins. The positive $\rho(H//b)$ at low fields can be understood by taking account of an enhancement of spin-disorder scattering as the antiferromagnetic state is changed into a field-induced ferromagnetic state. Since the magnetic scattering in a ferromagnetic state is much weaker than that in an antiferromagnetic state, however, the negative $\rho(H//b)$ at high fields could be a result of the strong reduction of scattering by the ferromagnetic alignment of Ce magnetic moments. For $H\perp b$ (the right panels of Fig. 8), the slope of magnetoresistance at 2 K changes the sign from positive to negative at higher magnetic fields, 14 and 8 T for $\rho_{//b}(H\perp b)$ and $\rho_{\perp b}(H\perp b)$, respectively. This result is compared to the slope change for $H//b$ at low magnetic fields, reflecting that the magnetic moments are easily aligned along the $b$ axis. In Fig. 8, it is noteworthy that $\rho_{\perp b}(H//b)$ at temperatures above $T_N^I$ appears to have a broad peak, which shifts to higher magnetic fields with increasing temperature. Both the larger negative magnetoresistance for $H//b$ and the slope change of magnetoresistance at lower field for $H//b$ are indicative of the strong ferromagnetic correlation along the $b$ axis. However, one should note that the negative magnetoresistance in CeNiGe$_2$ does not obey the scaling function, $\Delta\rho/\rho_o \sim -M^2$, expected for the conventional magnetoresistance due to the reduced scattering by field alignment of local spins. We can thus ascribe the negative magnetoresistance to the coexistence of magnetic correlation and heavy-fermion behavior and/or the Berry phase.

The combined results of $C(T, H)$ and $\rho(T, H)$ allow us to draw a $H$-$T$ phase diagram of CeNiGe$_2$, as shown in Fig. 9. In zero field, there are two magnetic transitions at $T_N^I = 4$ K and $T_N^{II} = 3$ K. With increasing magnetic field for both $H//b$ and $H\perp b$, the two magnetic transitions are combined into one. The value of magnetic field where the two transitions are combined into one is much lower when the field is applied along the $b$ axis, the easy magnetization direction.

## IV. CONCLUSIONS

The present studies on single crystals of CeNiGe$_2$ show a number of remarkable features. First, there is a very strong anisotropy in the magnetic and transport properties. The 2K magnetization ratio is observed to be 18 at 4 T and the zero-field resistivity ratio is about 70 at 300 K. This strong anisotropy is an unusual feature in lanthanide intermetallic compounds. The room temperature resistivity of CeNiGe$_2$ is more than 41 times for the isostructural but nonmagnetic CeNiSi$_2$.[16] The highly anisotropic properties of CeNiGe$_2$ are likely to result from different types of hybridization between the localized 4$f$ electrons and the conduction electrons and/or anisotropic magnetic ordering between the $b$ axis and in the plane. Second, two magnetic phase transitions observed at $T_N^I$ = 4 K and $T_N^{II}$ = 3 K have a little different nature in the antiferromagnetic orderings. The low-temperature data of magnetic susceptibility, specific heat, electrical resistivity, and Hall coefficient show two anomalies at $T_N^I$ and $T_N^{II}$. We suggest that the magnetic moments in CeNiGe$_2$ are strongly aligned along the $b$ axis antiferromagnetically, whereas the moments in the plane form an incommensurate structure in the antiferromagnetic ordering. This anisotropic hybridization could come the strong anisotropy in the magnetic and transport properties of CeNiGe$_2$. Third, the two magnetic states are significantly changed by the application of magnetic field. Field-dependent specific heat and magnetoresistance measurements indicate that the two magnetic transitions are combined to form one transition both for the field parallel and perpendicular to the $b$ axis. When the field is applied along the $b$ axis, the field value where the two transitions merge into one is much lower. In addition, the magnetic field parallel to the *b* axis induces a metamagnetic-like transitions from the antiferromagnetic state to a field-induced ferromagnetic state at very low field ~0.7 T.

Considering the substitution effect from the typical semimetallic silicon to the more metallic germanium, we believe that the substitution causes a remarkable increase in the density of states at the Fermi level and therefore in the electronic contribution to the specific heat. The electronic Sommerfeld coefficient $\gamma$ is enhanced from 45 mJ/K$^2$mol for CeNiSi$_2$ to 108 mJ/K$^2$mol for CeNiGe$_2$. The nonmagnetic ground state of CeNiSi$_2$ is also changed into the antiferromagnetic state in CeNiGe$_2$. The appearance of magnetic state on substituting germanium for silicon can be attributed to a strong hybridization between the localized 4*f* electrons and the conduction electrons due to the increase in the density of states at the Fermi level, and to a weakening of the interatomic interactions due to both the increase of the interatomic distance and a change in the chemical nature.

The Hall effect of CeNiGe$_2$ is quite unexpected, if considering the conventional skew scattering theory. Experimentally, there are some materials where an anomalous Hall effect was observed. For example, in colossal magnetoresistance manganites it was explained by Berry phase theory based on carrier hopping in a nontrivial spin background.[14] There occurs a very anomalous Hall effect in a sememetal CeSb,[17] where the presence of two types of carriers makes it difficult to analyze the Hall data. We expect that in CeNiGe$_2$, the anisotropic hybridization can, in the presence of spin-orbit coupling, give rise to the anomalous Hall effect.

The highly anisotropic magnetoresistance is the most striking feature in CeNiGe$_2$. If we compare the low-temperature ($T < 4$ K) data for *H*//*b* to those for *H*⊥*b*, the magnetic field where the magnetoresistance makes a maximum is much lower than for *H*⊥*b*. This result is consistent with the magnetization measurements, which display two characteristic fields changing the slope. For $H//b$, one is about 0.7 T where a maximum occurs both in d$M$/d$H$ and $\Delta\rho_{//b}/\rho_o$ and the other is about 3 T where $M(H)$ starts to saturate and $\Delta\rho_{\perp b}/\rho_o$ has a maximum. Hence, the change of magnetoresistance for $H//b$ could be attributed to competition between the enhancement of spin-disorder scattering from the antiferromagnetic to ferromagnetic orderings and the reduction of the scattering in the field-induced ferromagnetic state. For $H\perp b$, the two characteristic fields are observed at 8 and 14 T in $\rho_{\perp b}(H)$ and $\rho_{//b}(H)$, respectively. This result suggests that a higher magnetic field is required to align the magnetic moments perpendicular to the $b$ axis, leading to a retarded maximum for $H\perp b$. In paramagnetic region ($T > 4$ K), it is found that $\Delta\rho_{//b}/\rho_o$ exhibits a broad peak at high fields, while $\Delta\rho_{\perp b}/\rho_o$ is negative. This anisotropy is likely to be associated with scattering relaxation rates influenced by anisotropic crystalline electric field arising from the quasi-2D crystal structure.

**Acknowledgement**

We thank Prof. H. Nakotte and Dr. K. H. Ahn for their useful discussions. This work was supported by the National Science Foundation, the US Department of Energy, and the State of Florida. One of us (MHJ) acknowledges partial support from LANSCE - LANL.

**Figure captions**

Fig. 1. Isothermal magnetization $M(H)$ versus applied field to 50 T for $H//b$ and $H\perp b$ at 0.5 K for a single crystal of CeNiGe$_2$. The inset shows the low-field data of $M(H)$ at 2 K. The arrow indicates the metamagnetic-like transition field $H_m$ along the $b$ axis.

Fig. 2. Inverse of magnetic susceptibility $\chi^{-1}(T)$ versus temperature between 2 and 350K for CeNiGe$_2$ in an applied magnetic field of 0.1T for $H//b$ and $H\perp b$. The insets show the low-temperature data of $\chi_{//b}(T)$ and $\chi_{\perp b}(T)$.

Fig. 3. In-plane resistivity $\rho_{\perp b}(T)$ and $b$-axis resistivity $\rho_{//b}(T)$ for CeNiGe$_2$ as a function of temperature. The inset shows $\rho_{\perp b}(T)$ in the logarithmic scale of the temperature.

Fig. 4. Low-temperature data of (a) specific heat $C(T)$, (b) temperature derivative of in-plane resistivity $d\rho_{\perp b}/dT$, and (c) Hall coefficient $R_H(T)$ for CeNiGe$_2$.

Fig. 5. Isothermal magnetization $M(H)$ versus applied field to 10 T for CeNiGe$_2$ at several temperatures 0.5, 3.5, 5, and 20 K for $H//b$ and at 0.5 K for $H\perp b$.

Fig. 6. Specific heat $C(T)$ versus temperature at fixed magnetic fields, zero, 2, 4, and 6 T for $H\perp b$ and $H//b$.

Fig. 7. Electrical resistivity $\rho(T)$ versus temperature between 2 and 20K for CeNiGe$_2$ in various fields 0, 2, 5, 10, 15, 18T for $H//b$ and $H\perp b$.

Fig. 8. Normalized magnetoresistance $\Delta\rho/\rho_o = [\rho(H)-\rho(0)]/\rho(0)$ for CeNiGe$_2$ as a function of applied magnetic field for $H//b$ and $H\perp b$ at fixed temperatures.

Fig. 9. Magnetic phase diagram ($H$-$T$ diagram for $H//b$ and $H\perp b$). ○: $C(T, H//b)$, ●: $C(T, H\perp b)$, □: $\rho(T, H//b)$, and ■: $\rho(T, H\perp b)$.

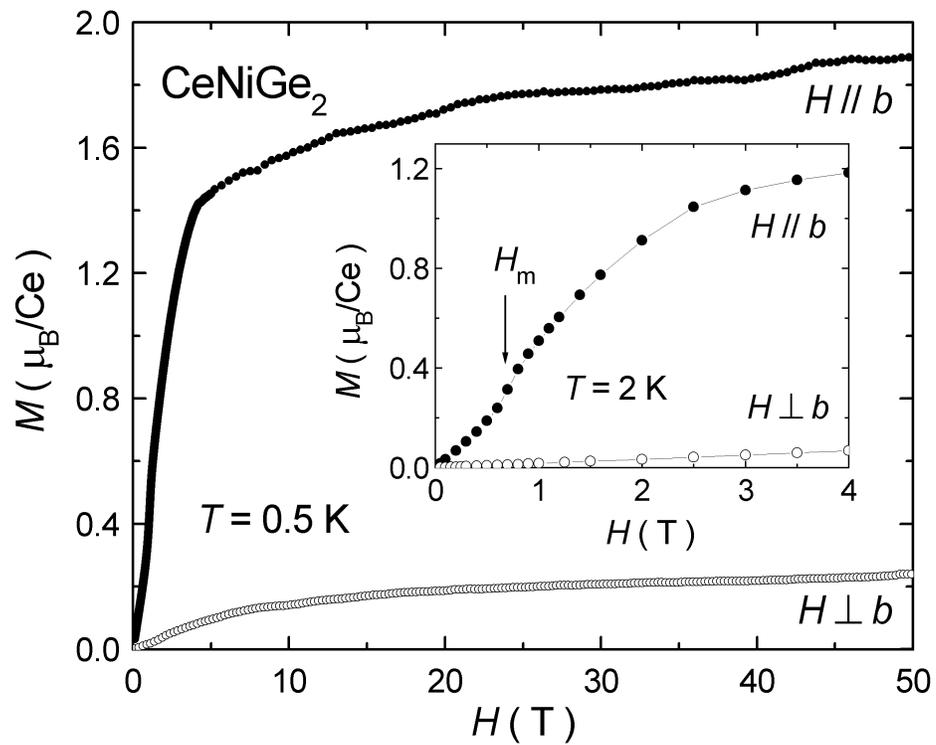

Fig. 1. M. H. Jung et al.

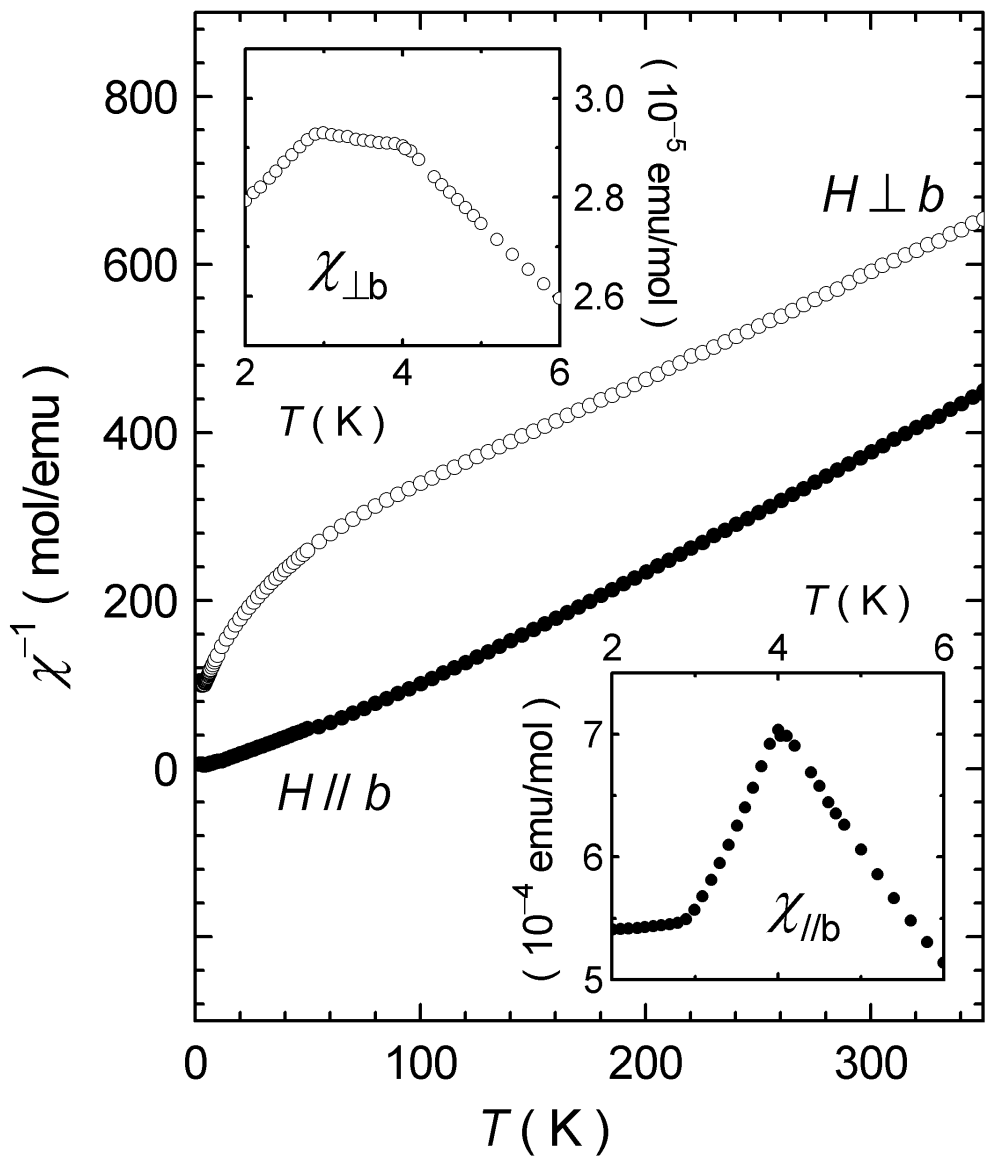

Fig. 2. M. H. Jung et al.

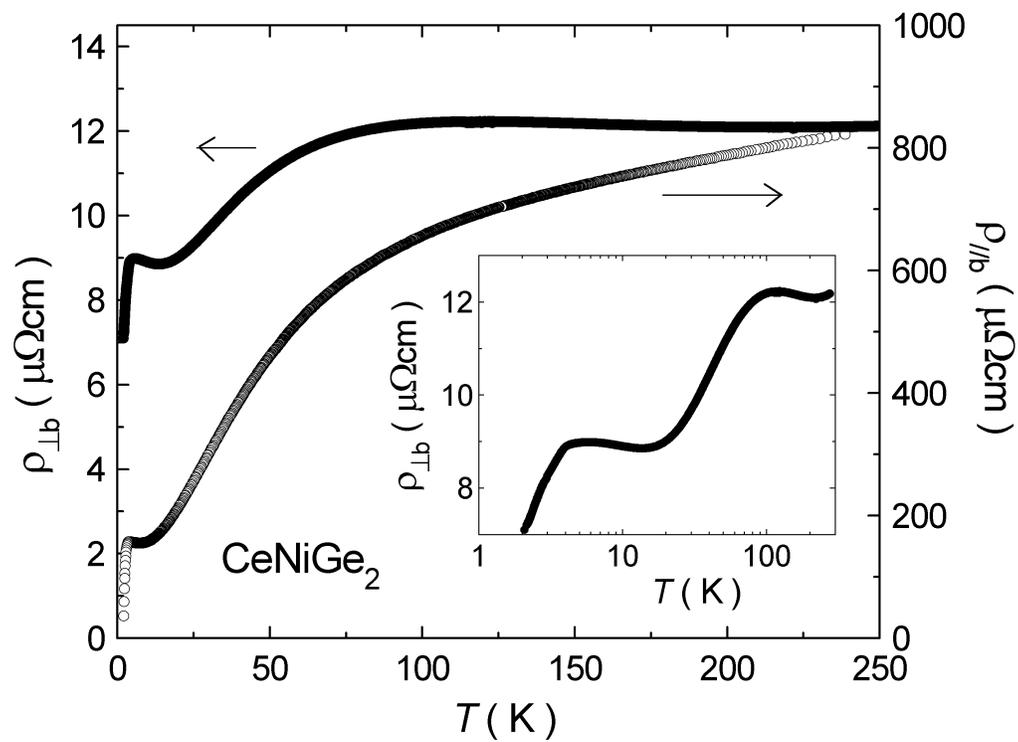

Fig. 3. M. H. Jung et al.

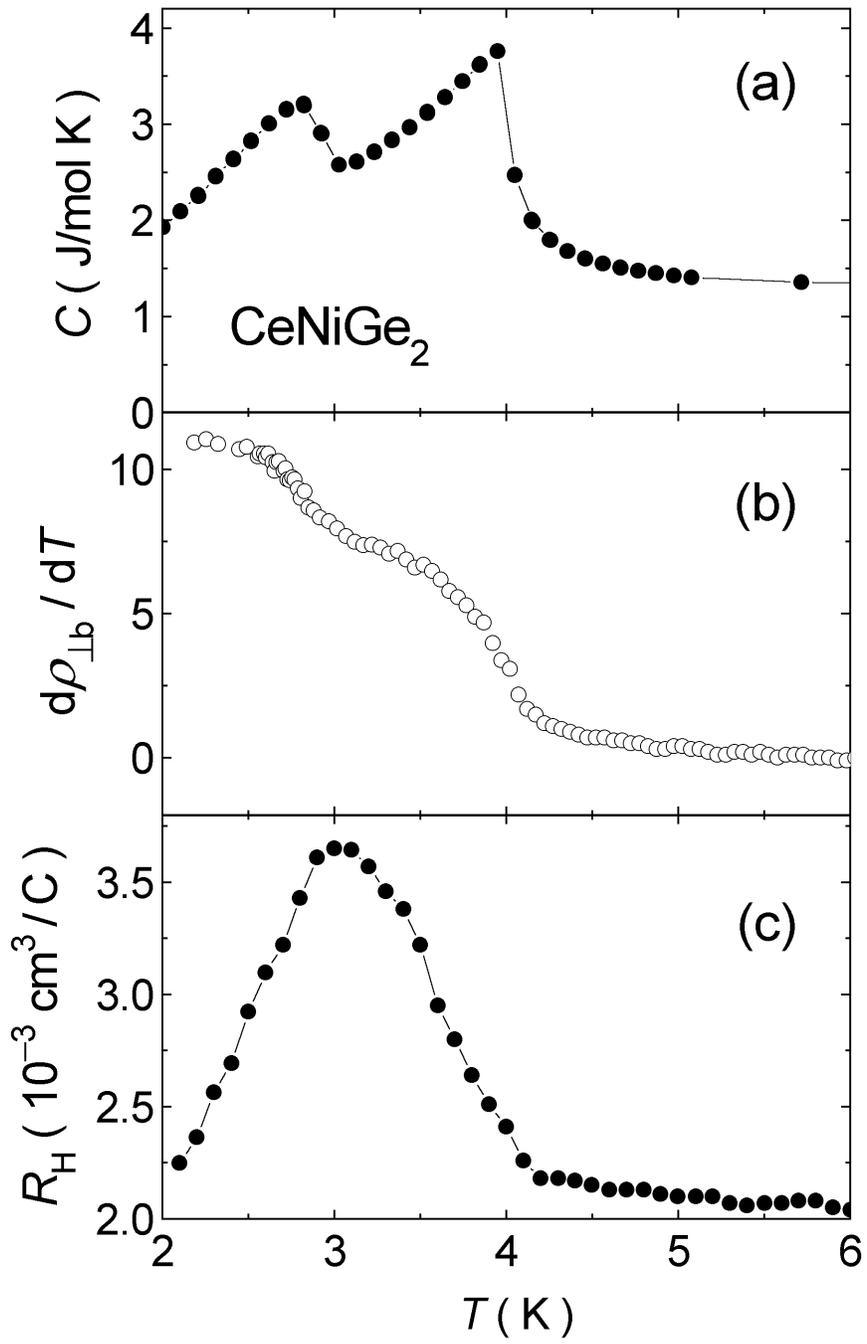

Fig. 4. M. H. Jung et al.

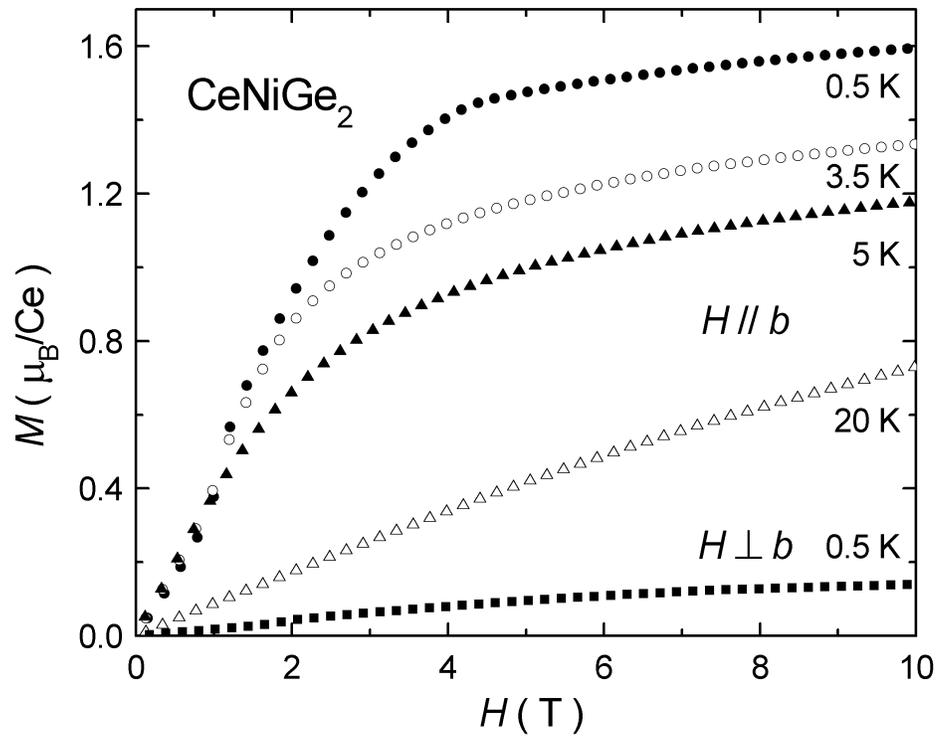

Fig. 5. M. H. Jung et al.

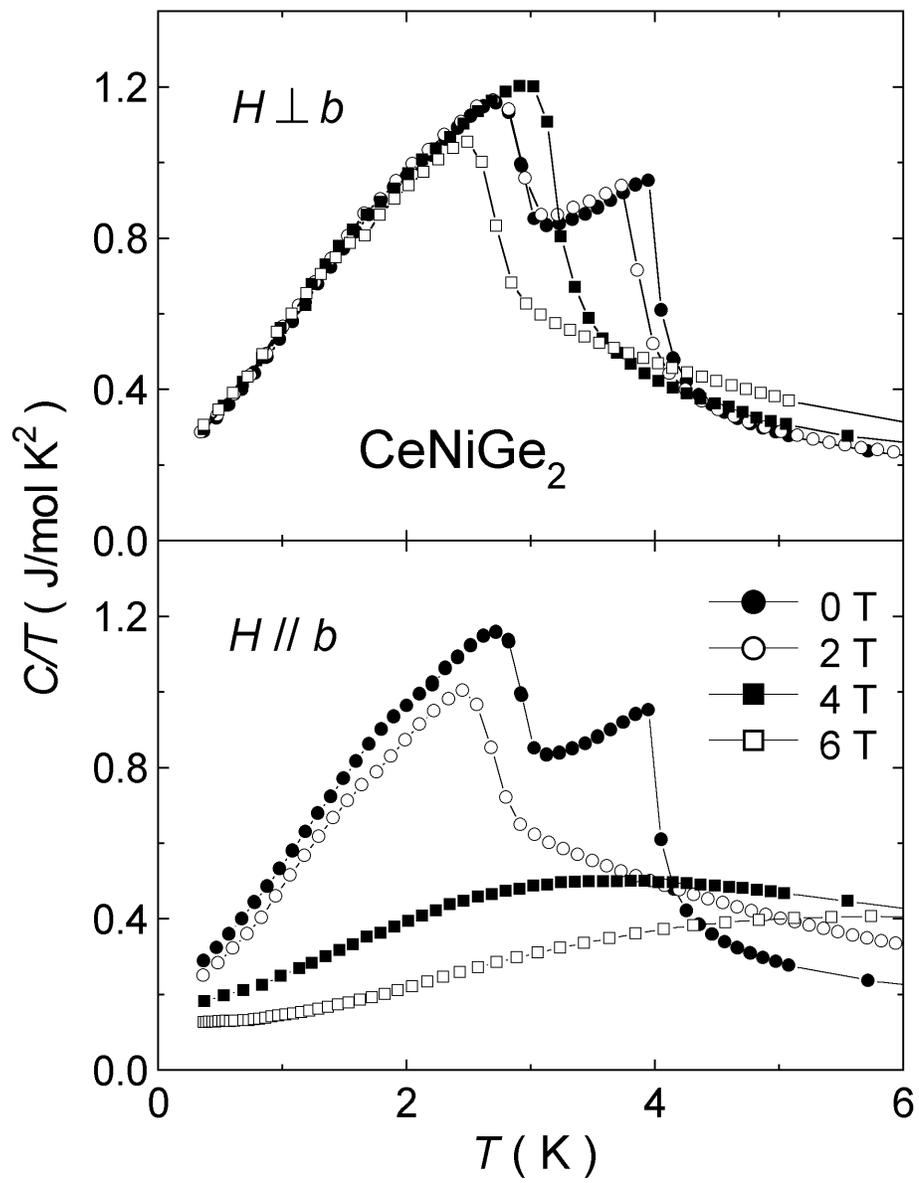

Fig. 6. M. H. Jung et al.

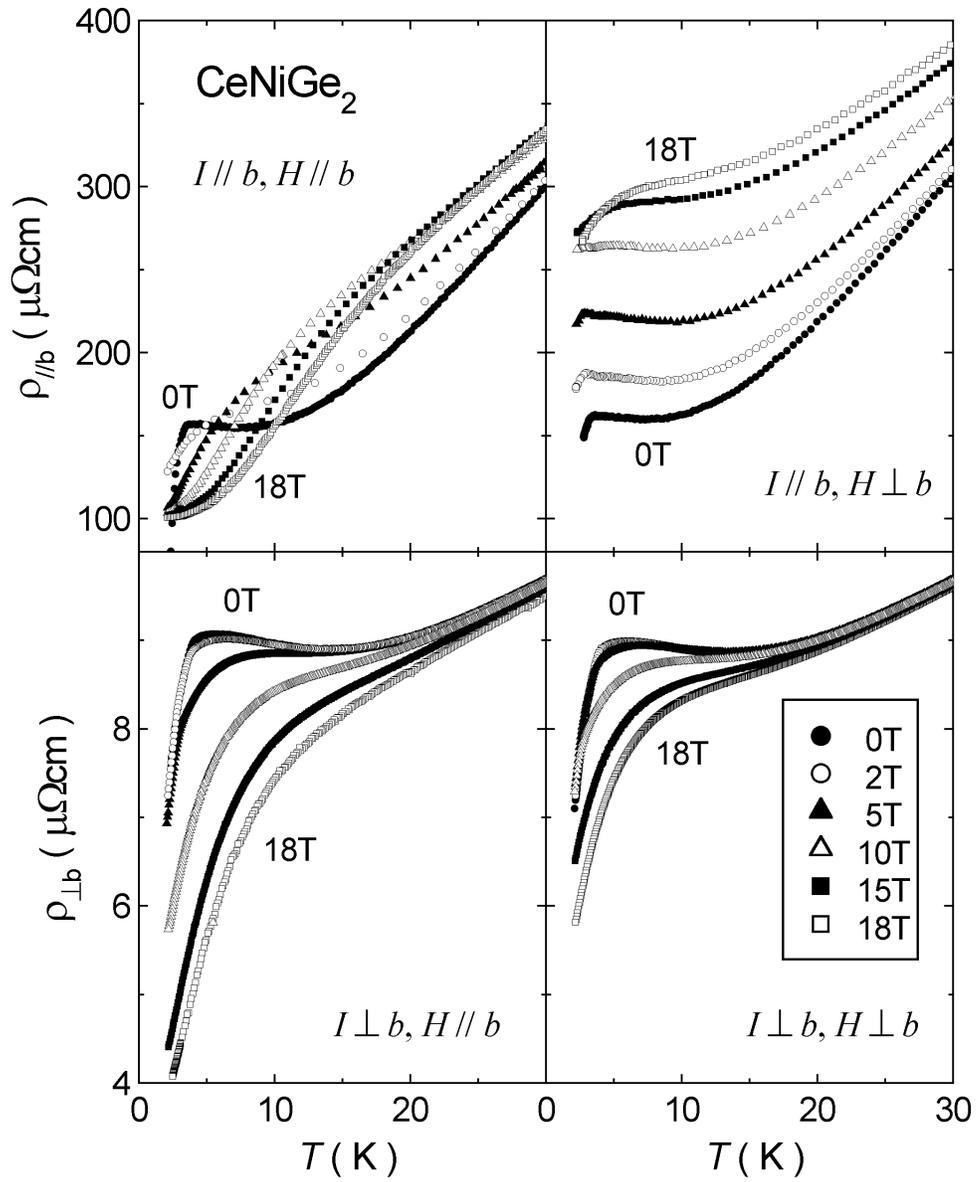

Fig. 7. M. H. Jung et al.

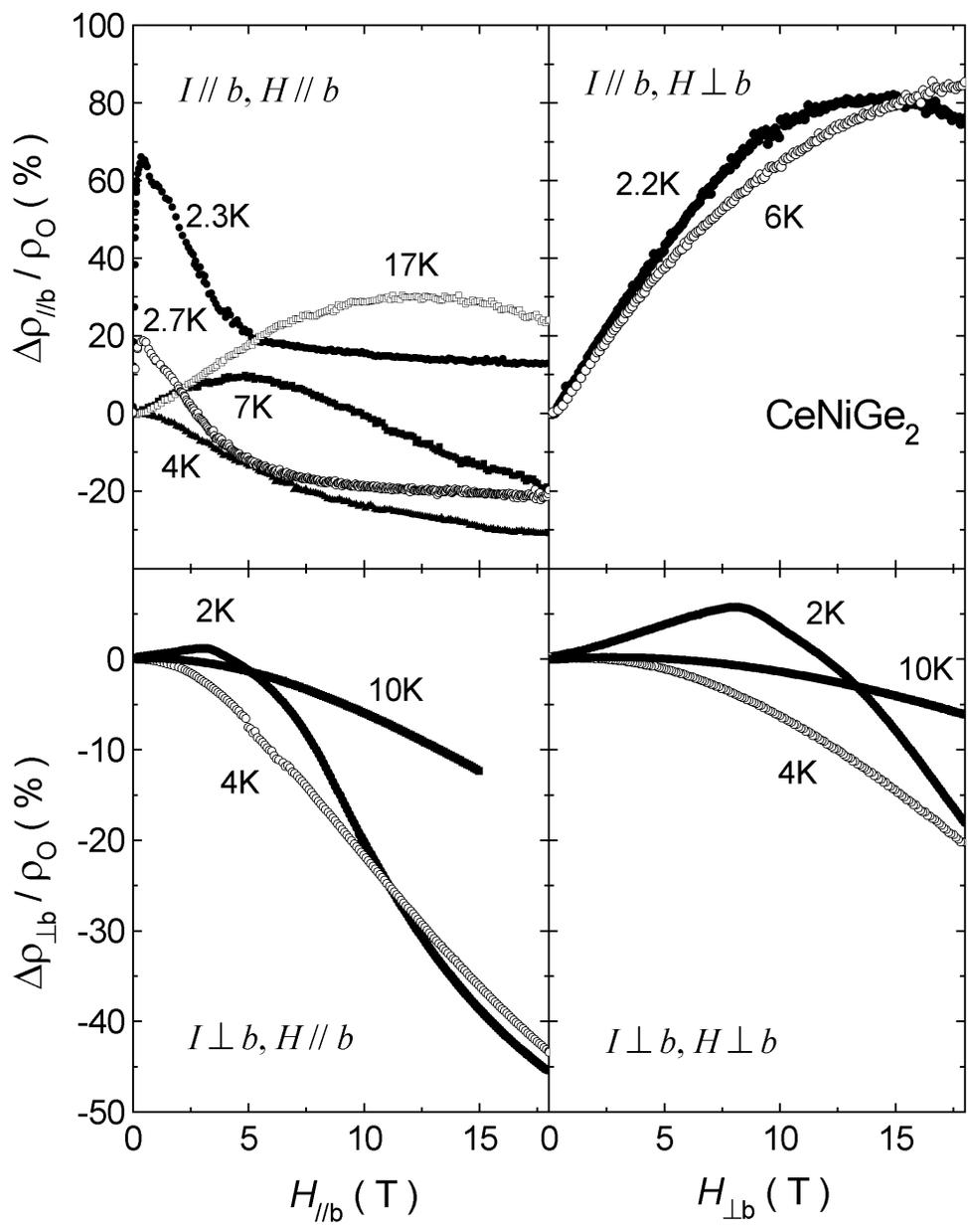

Fig. 8. M. H. Jung et al.

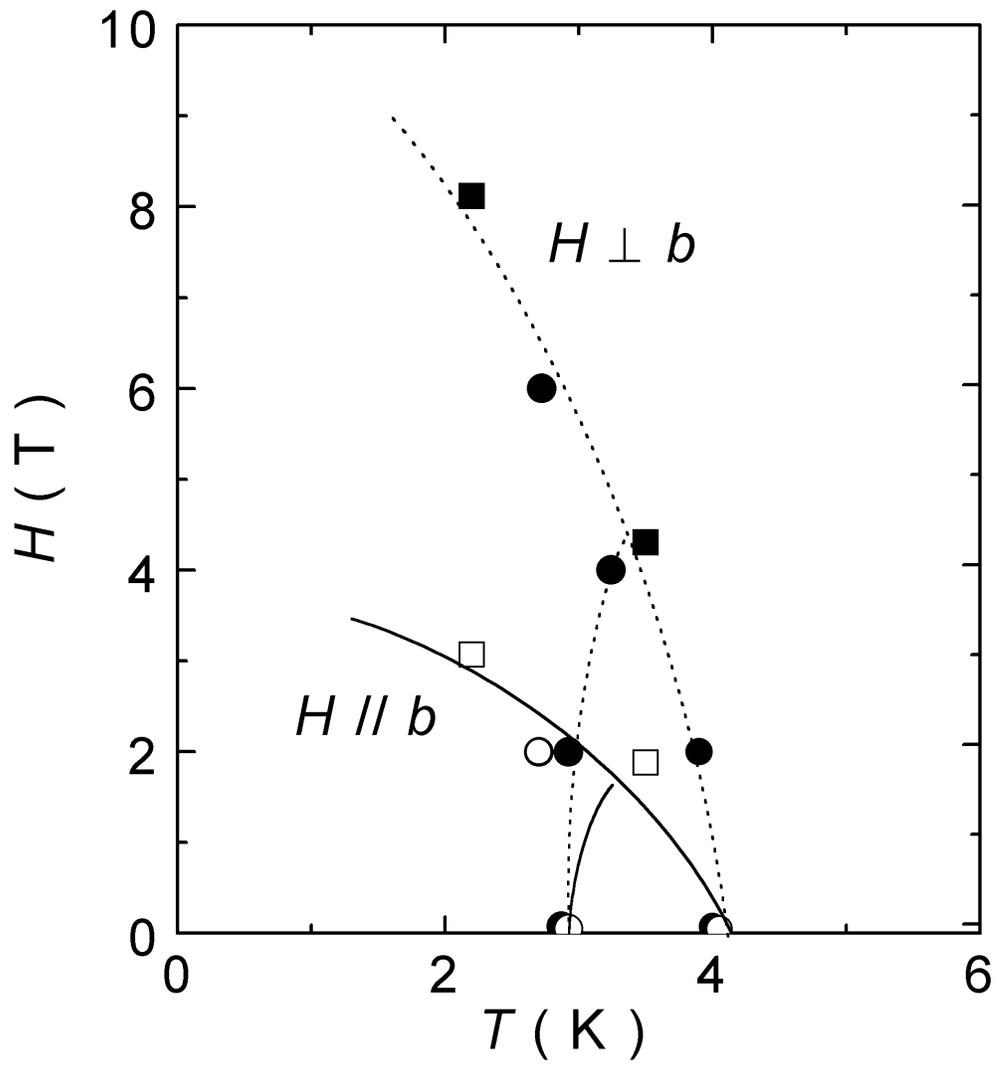

Fig. 9. M. H. Jung et al.